\documentclass[aps,pre,superscriptaddress,12pt]{revtex4}%
\usepackage{amssymb}%
\usepackage{amsmath}%
\usepackage{amsfonts}%
\usepackage{graphicx}
\begin{document}

\title{\bf Turing patterns in two dimensional reaction-diffusion system: effect of an electric field}

\author{B K Agarwalla \footnote{bijay.physics@gmail.com}}
\affiliation{Department of Physics, IIT Bombay, Mumbai 400076, India}

\author{J K Bhattacharjee \footnote{jkb@bose.res.in}}
\affiliation{S.N.Bose National Centre for Basic Sciences, JD Block, Sector III, Salt lake city, Kolkata 700098, India}

\author{P Titum }
\affiliation{Department of Physics, IIT Kanpur, 208016, India}

\begin{abstract}
  We consider a two dimensional Turing like system with two diffusing species which interact with each other. Considering the species to be charged, we include the effect of an electric field along a given direction which can lead to a drift induced instability found by A.B.Rovinsky and M.Menzinger\cite{9}.  This allows one to study the competition between diffusion and drift as was done numerically by Riaz et al. We show here that an analytic formula can be found on the basis of a linear stability analysis that incorporates all the effects that are known for the system and also allows for some detailed predictions.

\end{abstract}

\maketitle

Pattern formation in two reacting and diffusing species was first studied by Turing\cite{1,2}. Turing argued that if the diffusion coefficients of the two species are widely different, then if one of the species is auto catalytic with the other inhibiting its growth, then the steady homogeneous state will be unstable to a patterned steady state.  The instability could also set in as a temporal pattern in a spatially homogeneous state under certain conditions. Turing patterns have been a very important aspect of the study of nonlinear systems\cite{3,4,5,6,7,8}. Decades later it was found by Rovinsky and Menzinger \cite{9,10,11} that pattern formation could occur even if the two diffusion coefficients  were nearly equal provided there was an external electric field and the diffusion species were charged giving rise to a gradient coupling. The pattern formed in this case would be travelling waves as opposed to the standing waves of the Turing pattern. Recently there has been extension of the Turing work in some unexpected directions \cite{12,13}. In a recent study, Riaz et al\cite{14,15} have shown numerically that a Turing pattern for charged species could be altered by an applied electric field \cite{16}.

 In this work, we revisit the pattern formation problem with an external electric field to arrive at a single analytic result that has the property of capturing all the possible conditions for the instabilities in the system. We work with a two dimensional set up as in the work of Riaz et al. We take the system to be unbounded in the $y$ direction and  be bounded by rigid plates at $x=\pm L$. The boundary conditions are that the concentration of the species vanishes at the boundary and so does the current normal to the plates which is proportional to the $x$-derivative of the concentrations.  The electric field is taken to be in the $x$ direction which leads to a drift in that direction. The existence of the plates in the $x$-direction is vital to keep the system bounded. The plates also play the very important role of fixing the wave number in the $x$-direction. In the absence of the constraint there will be an overall selection mechanism for the wave numbers (${\bf k=k_{1}+k_{2}}$) but the individual components are not uniquely determined. What we will see below is that $k_{1}$ is fixed by the boundary condition and thus once $k^{2}$ is known both $k_{1} $and $k_{2}$ will be determined. In the work of Riaz et al\cite{14} the fixing of $ k_{1}$ and $k_{2}$ is undertaken numerically. Here the analytic  fixing of $ k_{1}$ allows us to find a general expression for the thresholds of the different instabilities.

The general reaction-diffusion problem for two species $A(x,y,t)$ and $B(x,y,t)$ can be modelled by the evalutaion equation\cite{14}

\begin{eqnarray}
\frac{\partial A}{\partial t} &=& D {\nabla^{2} A}+z_{2} E D \frac{\partial A}{\partial x}+f(A,B)\\
\frac{\partial B}{\partial t} &=& {\nabla^{2} B}+z_{1} E \frac{\partial B}{\partial x}+g(A,B)
\end{eqnarray}

In the above, $D$ is the diffusion coefficient for the species $A$ in units where the diffusion coefficient for the species $B$ is unity. The external electric field in the $x$-direction in denoted by $E$ and $z_{1}$ and $z_{2}$ are the charges associated with the inhibitor and the activitor respectively. The operator $\nabla^{2}$ is two dimensional (pattern on a substrate) and the function $f(A,B)$ and $g(A,B)$ describe the growth and interaction of the species $A$ and $B$. The electric field terms come from an expression for the current together with the relevant Einstein's relation. In the Gierer-Meinhardt model\cite{17,18},

\begin{eqnarray}
f(A,B)&=& \frac{A^{2}}{B}-A+\sigma\\
g(A,B)&=& \mu (A^{2}-B)
\end {eqnarray}
The growth rate of $A$ due to interaction with the substrate is $\sigma$ and the natural decay rate for $B$ is $\mu$. In the Lengyel-Epstein model\cite{19}
\begin{eqnarray}
f(A,B)&=&\sigma b (B-\frac{A B}{1+B^{2}})\\
g(A,B)&=&a-B-\frac{4AB}{1+B^{2}}
\end{eqnarray}

where $\sigma$, $b$ and $a$ are constants. The homogeneous steady state is $A=A_{0}$ and $B=B_{0}$ such that $f(A_{0},B_{0})=g(A_{0},B_{0})=0$. The linear stability analysis around $A=A_{0}$ and $B=B_{0}$ leads to 
\begin{eqnarray}
\frac{d\delta A}{dt} &=&D \nabla^{2} \delta A +z_{2} E D \frac{\partial (\delta A)}{\partial x}+a_{11} \delta A +a_{12} \delta B\\
\frac{d\delta B}{dt} &=&  \nabla^{2} \delta B +z_{1} E  \frac{\partial(\delta B)}{\partial x}+a_{21} \delta A +a_{22} \delta B
\end{eqnarray}

where $a_{11}=\left(\frac{\partial f}{\partial A}\right)_{A_{0},B_{0}}$, $a_{12}=\left(\frac{\partial f}{\partial B}\right)_{A_{0},B_{0}}$, $a_{21}=\left(\frac{\partial g}{\partial A}\right)_{A_{0},B_{0}}$ and  $a_{22}=\left(\frac{\partial g}{\partial B}\right)_{A_{0},B_{0}}$
   
We consider a geometry which is confined by plates at $x=\pm L$ and is unbounded in the $y$-direction. The solution will be periodic in the $y$-direction and if we take the wavenumber in this direction to be $k_{2}$, then we can write
\begin{equation}
(\delta A,\delta B)=(A_{1}(x),B_{1}(x))\exp{(ik_{2}y)}\exp{(\lambda t)}
\end{equation}
where $\lambda$ is the eigenvalue determining the temporal growth. Then $A_{1}(x)$ and $B_{1}(x)$ satisfy the differential equation

\begin{equation}
[\lambda + D k_{2}^{2}-D \frac{\partial^{2}}{\partial x^{2}}-z_{2} E D \frac{\partial}{\partial x}-a_{11}]A_{1} =a_{12}B_{1}\\
\end{equation}
\begin{equation}
[\lambda + k_{2}^{2}-\frac{\partial^{2}}{\partial x^{2}}-z_{1} E D \frac{\partial}{\partial x}-a_{22}]B_{1} = a_{21}A_{1}
\end{equation}
Eliminating $B_{1}$ one can write

\begin{eqnarray}
&D&\frac{d^{4}A_{1}}{dx^{4}}+\overbrace{\alpha_{2}}\frac{d^{3}A_{1}}{dx^{3}}+[(Da_{22}+a_{11})-2Dk_{2}^{2}-(1+D)\lambda \nonumber \\
&+&z_{1}z_{2}DE^{2}]\frac{d^{2}A_{1}}{dx^{2}}+[\overbrace{\beta_{2}}-\lambda(z_{1}+z_{2}D)E]\frac{dA_{1}}{dx}+[\Delta+\lambda^{2}\nonumber \\
&-&\lambda \overbrace{\alpha_{1}}-k_{2}^{2}(a_{11}+Da_{22})+Dk_{2}^{4}]=0
\end{eqnarray}
where 
\begin{eqnarray}
\overbrace{\alpha_{2}}&=&ED(z_{1}+z_{2})\nonumber \\
\overbrace{\beta_{2}}&=&E[z_{1}a_{11}+z_{2}Da_{22}-(z_{1}+z_{2})Dk_{2}^{2}]\nonumber \\
\overbrace{\alpha_{1}}&=&a_{11}+a_{22}-(1+D)k_{2}^{2}\nonumber \\
\Delta&=& a_{11}a_{22}-a_{12}a_{21}
\end{eqnarray}

At this point the general procedure should be clear. We need to solve the homogeneous 4th order equation above. This will involve four arbitrary constants which have to be fixed by boundary conditions. Since the system is homogeneous the four conditions will lead to four homogeneous linear algebraic equations and for consistency the determinant has to vanish. The resulting equation fixes $\lambda$ in terms of $L$, $k_{2}$, $E$ and other system parameters. The requirement $ Re{\lambda}\geq 0$ for instability allows us to discuss the different situation that can occur. 
The above procedure in general and in principle cumbersome. We illustrate this in the simpler situation of $E=0$ and the Turing limit i.e. $D<<1$. The lesson that we learn here will be put to good use for the more complicated case. 

Accordingly, we set $E=0$ in Eq.(12) and obtain
\begin{eqnarray}
D\frac{d^{4}A_{1}}{dx^{4}}+[Da_{22}+a_{11}-2Dk_{2}^{2}-\lambda (1+D)]\frac{d^{2}A_{1}}{dx^{2}}\nonumber \\
+[\lambda^{2}-\overbrace{\alpha_{1}} \lambda+Dk_{2}^{4}-k_{2}^{2}(a_{11}+Da_{22})+\Delta]A_{1}=0
\end{eqnarray}
The general solution can be written as 
\begin{equation}
A_{1}=\sum_{i=1}^{4} c_{i}\exp {(jm_{i}x)}
\end{equation}
where $j=\sqrt{-1}$.Then from eq.(14) we have
\begin{eqnarray}
Dm^{4}-[Da_{22}+a_{11}-2Dk_{2}^{2}-\lambda(1+D)]m^{2}\nonumber \\
+[\lambda^{2}-\overbrace{\alpha_{1}} \lambda+Dk_{2}^{4}-k_{2}^{2}(a_{11}+a_{22}D)+\Delta]=0
\end{eqnarray}
For $D<<1$ (The Turing case) the two roots are approximately (we keep $Dk_{2}^{2}$ since $k_{2}$ is not known a-priori)
\begin{eqnarray}
m_{1}^{2}\simeq \frac{a_{11}-\lambda-2Dk_{2}^{2}}{D}\\
m_{2}^{2}\simeq \frac{\lambda^{2}-\alpha \lambda-k_{2}^{2}a_{11}+Dk_{2}^{4}+\Delta}{a_{11}-\lambda-2Dk_{2}^{2}}
\end{eqnarray}
where $\alpha=a_{11}+a_{22}-k_{2}^{2}$.

For even solutions  we can write
\begin{equation}
A_{1}=c_{1}\cos (m_{1}x)+c_{2}\cos (m_{2}x)
\end{equation}
Imposing the boundary conditions $A_{1}=0$ and no flux condition $\frac{dA_{1}}{dx}=0$ on $x=\pm L$ one can write
\begin{equation}
m_{1}\tan (m_{1}L)=m_{2}\tan (m_{2}L)
\end{equation}
In the limit $D<<1$, $m_{1}\rightarrow \infty$ and this require $m_{2}L \rightarrow \frac{\pi}{2}$ leads to the condition
\begin{eqnarray}
&\lambda^{2}&-[Tr A-(k_{2}^{2}+\frac{\pi^{2}}{4L^{2}})]\lambda-(k_{2}^{2}+\frac{\pi^{2}}{4L^{2}})a_{11}\nonumber \\
&+&\Delta+Dk_{2}^{4}+2Dk_{2}^{2}\frac{\pi^{2}}{4L^{2}}=0
\end{eqnarray}
where $Tr A=a_{11}+a_{22}$.

We note that $k_{2}^{2}+\frac{\pi^{2}}{4L^{2}}$ enters as a combination which we call $k^{2}$ and to $0(D)$. we can add to eq.(21) a term $D(\frac{\pi^{2}}{4L^{2}})^{2}$ without committing any significant error since $L$ is very large as well. In that case eq.(21) becomes 
\begin{equation}
\lambda^{2}-(Tr A- k^{2})\lambda-a_{11}k^{2}+Dk^{4}+\Delta=0
\end{equation}
This reproduces the Turing condition to the leading order in $D$, since we find the condition for instability is
\begin{equation}
\Delta-\frac{a_{11}^{2}}{4D}<0
\end{equation}
and the characteristic wave number $k_{min}$ is given by
\begin{equation}
k_{min}^{2}=\frac{a_{11}}{2D}
\end{equation}
The lesson that we learn from the above exercise is that the operator $\frac{d}{dx}$ can be effectively replaced by $i\frac{\pi}{2L}$ and using eq.(12), we determine $\lambda$ from
\begin{eqnarray}
\lambda^{2}&-&\lambda[Tr A-(1+D)k^{2}+\frac{i \pi}{2L}E(z_{1}+z_{2}D)]\nonumber \\
&+&Dk^{4}-k^{2}(a_{11}+a_{22}D)+\Delta-\frac{i \pi}{2L}EDz_{2}(k^{2}-a_{22})\nonumber \\
&-& \frac{i \pi}{2L}Ez_{1}(Dk^{2}-a_{11})-\frac{\pi^{2}}{4L^{2}}z_{1}z_{2}E^{2}D=0
\end{eqnarray} 
The above equation can be written as
\begin{equation}
\lambda^{2}-\lambda(\alpha_{1}+i\alpha_{2})+\beta_{1}+i\beta_{2}=0
\end{equation}
where 
\begin{eqnarray}
\alpha_{1}&=& Tr A-(1+D)k^{2}\nonumber \\
\alpha_{2}&=& \frac{\pi}{2L}E(z_{1}+z_{2}D)\nonumber \\
\beta_{1}&=& Dk^{4}-k^{2}(a_{11}+Da_{22})+\Delta-\frac{\pi^{2}}{4L^{2}}E^{2}z_{1}z_{2}D\nonumber\\
\beta_{2}&=&\frac{\pi}{2L}E[z_{1}a_{11}+z_{2}Da_{22}-(z_{1}+z_{2})Dk^{2}]
\end{eqnarray}
The real part of the eigenvalue $\lambda$ in eq.(25) will be negative (the condition that the homogeneous state will be stable) provided
\begin{equation}
\beta_{2}^{2}<\alpha_{1}(\alpha_{1}\beta_{1}+\alpha_{2}\beta_{2})
\end{equation}
using $\alpha_{1}$, $\alpha_{2}$, $\beta_{1}$, $\beta_{2}$  given in eq.(27) leads after straightforward algebra to the central result
\begin{eqnarray}
(z_{1}-Dz_{2})^{2}[a_{11}a_{22}-k^{2}(a_{11}+Da_{22})+Dk^{4}]\frac{\pi^{2}}{4L^{2}}E^{2}\nonumber\\
>[Tr A-(1+D)k^{2}]^{2}[k^{2}(a_{11}+Da_{22})-\Delta-Dk^{4}]
\end{eqnarray}
This is the single formula that contains all the possibilites of the pattern formation in the presence of the electric field.

We will now discuss the various possibilities.

First consider the Turing problem $E=0$. Now, the right hand side of eq.(29) has to be negative for stability which means
\begin{equation}
\Delta-k^{2}(a_{11}+Da_{22})+Dk^{4}>0
\end{equation}
for stability. This is the central criterion for stability for all $\Delta>0$ in the presence of diffusion. If the sign is reversed in eq.(30), we get the Turing instability and for $\Delta>0$, this occurs for a band of wavenumber where the above expression is negative. The minimum of the expression is obtained for $k^{2}=\frac{a_{11}+Da_{22}}{2D}$ and the value at the minimum is $\Delta-\frac{(a_{11}+Da_{22})^{2}}{4D}$ and hence the instability criterion is $\Delta<\frac{(a_{11}+Da_{22})^{2}}{4D}$ , an inequality which is easy to satisfy for $D<<1$

We now consider the opposite limit i.e. there is no diffusion and only drift. In this case eq.(29) acquires the form 
\begin{equation}
(z_{1}-Dz_{2})^{2}a_{11}a_{22}\frac{\pi^{2}}{4L^{2}}E^{2}>(Tr A)^{2}(-\Delta)
\end{equation}
Since we want to start with an initially stable state i.e. $Tr A<0$ and $\Delta>0$, we have $a_{11}a_{22}<0$ and eq.(31) becomes 
\begin{equation}
\frac{\pi E}{2L}<\frac{| Tr A|}{|z_{1}-Dz_{2}|}\left(\frac{-\Delta}{a_{11}a_{22}}\right)^{\frac{1}{2}}
\end{equation}
This clearly shows that an instability will set in if $E>E_{0}$, where
\begin{equation} 
\frac{\pi E_{0}}{2L}=\frac{| Tr A|}{|z_{1}-Dz_{2}|}\left(\frac{-\Delta}{a_{11}a_{22}}\right)^{\frac{1}{2}}
\end{equation}
We see immediately that for the instability to set in one must have a differential mobility i.e. $z_{1}\neq Dz_{2}$. This result in accordance with A.B.Rovinsky and M.Menzinger \cite{9}. In the case of $D\simeq 1$, i.e. the two diffusivites are nearly equal( a situation very different from Turing), we get for instability 
\begin{eqnarray}
(z_{1}-z_{2})^{2}[a_{11}a_{22}-k^{2}(a_{11}+a_{22})+k^{4}]\frac{\pi^{2}}{4L^{2}}E^{2}\nonumber \\
> (Tr A-2k^{2})^{2}[k^{2}(a_{11}+a_{22})-\Delta-k^{4}]
\end{eqnarray}

We treat the situation which for $E=0$ is stable so far as the reaction goes and is also stable when diffusion is included. This implies $\Delta>0$, $Tr A<0$
, $\Delta-(a_{11}+a_{22})k^{2}+k^{4}>0$. The right hand side of eq.(34) is now negative and for the inequality to hold, we need $a_{11}a_{22}-k^{2}(a_{11}+a_{22})+k^{4}<0$. With $a_{11}a_{22}<0$ and $Tr A<0$, we can satisfy this inequality in the range $0\leq k\leq k_{0}$ where 
\begin{equation}
2k_{0}^{2}=Tr A+\sqrt {(Tr A)^{2}-4a_{11}a_{22}}
\end{equation}
and for $k<k_{0}$, the critical value of $E$ which will trigger an instability will be given by 
\begin{equation}
\frac{\pi^{2}}{4L^{2}}E^{2}>\frac{(Tr A-2k^{2})^{2}}{(z_{1}-z_{2})^{2}}\frac{k^{2}(a_{11}+a_{22})-\Delta-k^{4}}{a_{11}a_{22}-k^{2}(a_{11}+a_{22})+k^{4}}
\end{equation}

\begin{figure}
\includegraphics[scale=.32]{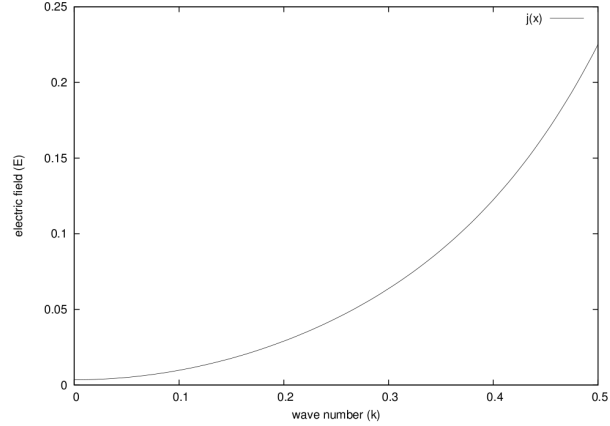} 
\caption{Plot of critical electric field ($E$) vs wave number ($k$) for $a_{11}=$0.899, $a_{22}=$-0.91, $a_{12}=$1, $a_{21}=$-0.899, $z_{1}=$1 and  $z_{2}=$2}
\label{figure1}
\end{figure}

According to the above relation all wavenumbers greater than $k_{0}$ are always stable. In writing down the above condition, we see the advantage of the exact expression of eq.(29). The order of magnitude estimation in \cite{9} does not yield the above answer.

For any $D$, we note that if diffusion destabilizes a stable reactive system, then $k^{2}(a_{11}+Da_{22})-\Delta-Dk^{4}>0$ and it follows that $a_{11}a_{22}-k^{2}(a_{11}+Da_{22})+Dk^{4}=a_{11}a_{22}-\Delta+[\Delta-k^{2}(a_{11}+Da_{22})+Dk^{4}]<0$ since $a_{11}a_{22}<0$. Hence eq.(29) can never be satisfied. No amount of electric field can stabilize the system. Finally, if the diffusive system is stable i.e. $k^{2}(a_{11}+Da_{22})-\Delta-Dk^{4}<0$, then a critical $E(E_{c})$ will destabilize the state provided $a_{11}a_{22}-k^{2}(a_{11}+Da_{22})+Dk^{4}<0$ which will happen if $k<k_{0}^{'}$ given by
\begin{equation}
k_{0}^{'2}=\frac{a_{11}+Da_{22}+\sqrt{(a_{11}+Da_{22})^{2}-4Da_{11}a_{22}}}{2D}
\end{equation}
For small enough $D$, $k_{0}^{'2}\simeq \frac{(a_{11}+Da_{22})}{D}$ and
\begin{equation}
\frac{\pi^{2}}{4L^{2}}E_{c}^{2}\simeq \frac{[Tr A-(1+D)k^{2}]^{2}}{(z_{1}-Dz_{2})^{2}} \left(\frac{-\Delta}{a_{11}a_{22}}\right)
\end{equation}

In summary we have studied a reaction-diffusion system in the presence of a constant electric field along a particular direction. We have found a single analytic expression which contains all possible information about the stability and the instability of the system for different ranges of the diffusion coeffecient. The primary result that emerges is that there is an uppar limit on the wave number of the instability and that for each wave number below that there is a critical electric field that can excite that particular wave number.

\end{document}